\documentclass[amsmath,amssymb,twocolumn]{revtex4}

\usepackage{graphicx}

\def\ri{{\rm i}}

\def\PT{$\mathcal{PT}$}

\begin{document}
\title{Partially-\PT-symmetric optical potentials with all-real spectra and soliton families in multi-dimensions}
\author{Jianke Yang}

\address{Department of Mathematics and Statistics, University of Vermont, Burlington, VT 05401, USA}

\begin{abstract}
Multi-dimensional complex optical potentials with partial
parity-time (\PT) symmetry are proposed. The usual \PT symmetry
requires that the potential is invariant under complex conjugation
and simultaneous reflection in all spatial directions. However, we
show that if the potential is only partially \PT-symmetric, i.e., it
is invariant under complex conjugation and reflection in a
\emph{single} spatial direction, then it can also possess all-real
spectra and continuous families of solitons. These results are
established analytically and corroborated numerically.
\end{abstract}

\maketitle

In optics, light propagation is often modeled by Schr\"odinger-type
equations \cite{Kivshar_book}. If the medium contains gain and loss,
the optical potential of the Schr\"odinger equation would be
complex. A surprising finding in recent years is that, if this
complex potential satisfies parity-time (\PT) symmetry, then the
linear spectrum can still be all-real, thus admitting stationary
light transmission
\cite{Bender1998,PT_2005,Christodoulides2007,Guo2009,Segev2010,PT_lattice_exp}.
Here \PT symmetry means that the potential is invariant under
complex conjugation and simultaneous reflection in all spatial
directions. In one dimension (1D), \PT-symmetry condition is
$V^*(x)=V(-x)$; in 2D, this condition is $V^*(x,y)=V(-x,-y)$; and so
on. Besides all-real spectra, \PT-symmetric potentials have been
found to support continuous families of optical solitons
\cite{Musslimani2008,Wang2011,Lu2011,Nixon2012, Zezyulin2012a}. But
if the complex potential is not \PT-symmetric, then the linear
spectrum is often non-real, and soliton families often do not exist
\cite{Yang_necessity}. Other findings on \PT systems can be found in
\cite{Abdullaev2011,coupler1,coupler2,He2012,Longhi_2009,Musslimani_diffraction_2010,Christodoulides_uni_2011,Nixon2012b,Nixon2013,Huang2013,Konotop2012,
Kevrekidis2013, Li2011, Zezyulin2012b,
Kartashov2013,Barashenkov2013}.

In this Letter, we show that in multi-dimensions, if the complex
potential is not \PT-symmetric but is partially-\PT-symmetric, then
such potentials can still admit all-real spectra and continuous
families of solitons. Here partial \PT-symmetry means that the
potential is invariant under complex conjugation and reflection in a
\emph{single} spatial direction (rather than in all spatial
directions simultaneously). For example, in 2D,
partially-\PT-symmetric potentials are such that either
$V^*(x,y)=V(-x,y)$ or $V^*(x,y)=V(x,-y)$. Partially-\PT-symmetric
potentials constitute another large class of complex potentials with
all-real spectra and soliton families, and they may find interesting
applications in optics. For simplicity, we consider the 2D case
throughout the Letter, but similar results hold for three and higher
dimensions too.

The model for nonlinear propagation of light beams in complex
optical potentials is taken as
\begin{equation} \label{Eq:NLS}
\ri \Psi_z + \nabla^2 \Psi + V(x,y)\Psi + \sigma |\Psi|^2 \Psi = 0,
\end{equation}
where $z$ is the propagation direction, $(x,y)$ is the transverse
plane, $\nabla^2=\partial_{xx}+\partial_{yy}$, and $\sigma=\pm 1$ is
the sign of nonlinearity. The complex potential $V(x,y)$ is assumed
to possess the partial \PT symmetry
\begin{equation} \label{e:PPTcondition}
V^*(x,y)=V(-x,y).
\end{equation}
The real part of this potential is symmetric in $x$, and its
imaginary part anti-symmetric in $x$. No symmetry is assumed in the
$y$ direction.

First, we show that the spectrum of this partially-\PT-symmetric
potential can be all-real. Eigenvalues of this potential are defined
by the Schr\"odinger equation
\begin{equation} \label{e:eigen}
(\nabla^2+V)\psi=\lambda\psi,
\end{equation}
where $\lambda$ is the eigenvalue and $\psi$ the eigenfunction.

We start by considering separable potentials, where
\begin{equation*}
V(x,y)=V_1(x)+V_2(y).
\end{equation*}
For these potentials, the partial \PT symmetry condition
(\ref{e:PPTcondition}) implies that
\begin{equation*}
V_1^*(x)=V_1(-x), \quad V_2^*(y)=V_2(y).
\end{equation*}
Thus the function $V_1(x)$ is \PT-symmetric and $V_2(y)$ strictly
real. Eigenvalues of this separable potential are
\begin{equation*}
\lambda=\Lambda_1+\Lambda_2,
\end{equation*}
and the corresponding eigenfunctions are
$\psi(x,y)=\Psi_1(x)\Psi_2(y)$, where
\begin{equation*}
\left[\partial_{xx}+V_1(x)\right]\Psi_1(x)=\Lambda_1\Psi_1(x),
\end{equation*}
\begin{equation*}
\left[\partial_{yy}+V_2(y)\right]\Psi_2(y)=\Lambda_2\Psi_2(y).
\end{equation*}
Since $V_1(x)$ is \PT-symmetric, its eigenvalues $\Lambda_1$ can be
all-real. Since $V_2(y)$ is strictly real, its eigenvalues
$\Lambda_2$ are all-real as well. Thus eigenvalues $\lambda$ of the
separable potential $V(x,y)$ can be all-real.

Next we consider separable potentials perturbed by localized
potentials,
\begin{equation} \label{e:Vpert}
V(x,y)=V_0(x,y)+\epsilon V_p(x,y),
\end{equation}
where $V_0$ is separable, $V_p$ localized, $\epsilon$ a small real
parameter, and both $V_0, V_p$ satisfy the partial-\PT-symmetry
condition~(\ref{e:PPTcondition}). Since $V_p$ is localized,
continuous eigenvalues of the perturbed potential $V$ are the same
as those of the separable potential $V_0$ and are thus all-real. We
now show that discrete eigenvalues of $V$ are also real.

Suppose $\lambda_0$ is a simple discrete real eigenvalue of the
separable potential $V_0$. Since $V_0$ is partially-\PT-symmetric,
the eigenfunction $\psi_0$ of $\lambda_0$ is partially-\PT-symmetric
as well, i.e., $\psi_0^*(x,y)=\psi_0(-x,y)$. Under perturbation
$\epsilon V_p$, the perturbed eigenvalue and eigenfunction can be
expanded into the following perturbation series,
\begin{equation*}
\lambda=\lambda_0+\epsilon \lambda_1+\epsilon^2\lambda_2+\dots,
\end{equation*}
\begin{equation*}
\psi=\psi_0+\epsilon \psi_1+\epsilon^2\psi_2+\dots.
\end{equation*}
Substituting these expansions and the perturbed potential
(\ref{e:Vpert}) into Eq. (\ref{e:eigen}), at $O(\epsilon)$ we get
\begin{equation} \label{e:Lpsi1}
L\psi_1=(\lambda_1-V_p)\psi_0,
\end{equation}
where $L\equiv \nabla^2+V_0-\lambda_0$. Since $\lambda_0$ is a
simple eigenvalue, the kernel of the adjoint operator $L^*$ then
contains a single eigenfunction $\psi_0^*$. Then in order for Eq.
(\ref{e:Lpsi1}) to be solvable, the solvability condition is that
its right hand side be orthogonal to $\psi_0^*$, which yields
\begin{equation} \label{f:lambda1}
\lambda_1=\frac{\langle \psi_0^*, V_p\psi_0\rangle}{\langle \psi_0^*, \psi_0\rangle},
\end{equation}
where the inner product is defined as
\begin{equation*}
\langle f, g \rangle=\int_{-\infty}^\infty \int_{-\infty}^\infty  f^*(x,y) g(x,y) dx dy.
\end{equation*}
Since $\lambda_0$ is simple, it is easy to show that $\langle
\psi_0^*, \psi_0\rangle\ne 0$.
%it then does not admit generalized eigenfunctions $\phi$ satisfying
%$L\phi=\psi_0$, hence the solvability condition of this $\phi$
%equation cannot be satisfied, meaning that $\langle \psi_0^*,
%\psi_0\rangle\ne 0$ in Eq. (\ref{f:lambda1}).

A key consequence of partial \PT symmetry is that, if functions $f$
and $g$ are both partially-\PT-symmetric, then their inner product
$\langle f, g\rangle$ is real, because
\begin{equation*}
\langle f, g\rangle^* = \langle f^*, g^*\rangle=\langle f(-x, y), g(-x, y)\rangle=\langle f, g\rangle.
\end{equation*}
Since $\psi_0$ and $V_p$ are partially-\PT-symmetric, the inner
products in Eq. (\ref{f:lambda1}) then are real, thus $\lambda_1$ is
real.

Pursuing this perturbation calculation to higher orders, we can show
that $\lambda_n$ is real for all $n\ge 1$, thus the eigenvalue
$\lambda$ remains real under perturbations $\epsilon V_p$.

For general partially-\PT-symmetric potentials, we use numerical
methods to establish that their spectra can be all-real. To
illustrate, we take the complex potential $V(x,y)$ to be
\begin{eqnarray} \label{e:Vexample}
V(x,y)=3\left(e^{-(x-x_0)^2-(y-y_0)^2}+e^{-(x+x_0)^2-(y-y_0)^2}\right) \nonumber \\ +2\left(e^{-(x-x_0)^2-(y+y_0)^2}+e^{-(x+x_0)^2-(y+y_0)^2}\right)   \nonumber \\
+i\beta \left[2\left(e^{-(x-x_0)^2-(y-y_0)^2}-e^{-(x+x_0)^2-(y-y_0)^2}\right) \right. \nonumber \\ \left. +\left(e^{-(x-x_0)^2-(y+y_0)^2}-e^{-(x+x_0)^2-(y+y_0)^2}\right)\right],
\end{eqnarray}
where we set $x_0=y_0=1.5$, and $\beta$ is a real constant. This
potential is not \PT-symmetric, but is partially-\PT-symmetric with
symmetry (\ref{e:PPTcondition}). For $\beta=0.1$, this potential is
displayed in Fig. 1 (top row). It is seen that Re($V$) is symmetric
in $x$, Im($V$) anti-symmetric in $x$, and both Re($V$), Im($V$) are
asymmetric in $y$. The spectrum of this potential is plotted in Fig.
1(c). It is seen that this spectrum contains three discrete
eigenvalues and the continuous spectrum, which are all-real. Thus we
have numerically established that partially-\PT-symmetric potentials
can have all-real spectra. For these real eigenvalues, their
eigenfunctions respect the partial \PT symmetry of the potential.

\begin{figure}
\begin{center}
\includegraphics[height = 3in]{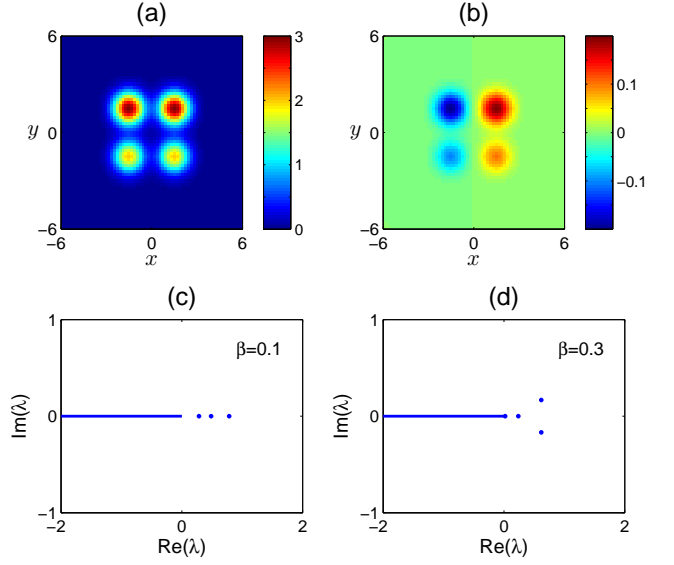}
\caption{(a,b) Real and imaginary parts of the partially-\PT-symmetric potential
(\ref{e:Vexample}) for $\beta=0.1$; (c, d) spectrum of this potential for $\beta=0.1$ and 0.3 respectively. }
\end{center}
\end{figure}

For potential (\ref{e:Vexample}) with varying $\beta$, we have found
that its spectrum is all-real as long as $|\beta|$ is below a
threshold value of $0.214$. Above this threshold, a phase transition
occurs, where complex eigenvalues appear in the spectrum, and their
eigenfunctions lose the partial \PT symmetry. This phase transition
is illustrated in Fig. 1(d), where the spectrum at $\beta=0.3$ is
shown. Phase transition is a well-known phenomenon of \PT-symmetric
potentials \cite{Bender1998,Guo2009,Segev2010,Musslimani2008}. We
see that it arises in partially-\PT-symmetric potentials too.

Next we examine whether these partially-\PT-symmetric potentials
support continuous families of solitons. These solitons are special
solutions of Eq. (\ref{Eq:NLS}) in the form of
\begin{equation} \label{e:solitonform}
\Psi(x,y,t)=\psi(x,y)e^{i\mu z},
\end{equation}
where $\mu$ is a real propagation constant, and $\psi(x,y)$
satisfies the equation
\begin{equation} \label{e:soliton}
\nabla^2 \psi + V(x,y)\psi + \sigma |\psi|^2 \psi=\mu \psi
\end{equation}
and vanishes when $(x,y)$ goes to infinity. In 1D, non-\PT-symmetric
potentials cannot admit soliton families \cite{Yang_necessity}.
However, in higher dimensions, we will show analytically and
numerically that partially-\PT-symmetric potentials do support
continuous families of solitons.

First, we show analytically that, from each real discrete eigenvalue
of the partially-\PT-symmetric potential, a continuous family of
solitons bifurcates out under each of the focusing and defocusing
nonlinearities. Suppose $\mu_0$ is a discrete simple real eigenvalue
of the potential and $\psi_0$ is its eigenfunction, i.e.,
$L\psi_0=0$, where $L\equiv \nabla^2+V-\mu_0$. Then we seek solitons
with the following perturbation expansion
\[
\psi(x,y; \mu) = \epsilon^{1/2} \left[c_0\psi_0 + \epsilon \psi_1 +\epsilon^2
\psi_2 +\dots\right],   \label{e:psiexpand}
\]
where $\epsilon \equiv |\mu-\mu_0|\ll 1$, and $c_0$ is a certain
non-zero constant. Substituting this expansion into Eq.
(\ref{e:soliton}), the $O(\epsilon^{1/2})$ equation is automatically
satisfied. At $O(\epsilon^{3/2})$, we get the equation for $\psi_1$
as
\[  \label{e:psi1}
L\psi_1=c_0\left(\rho \psi_0-\sigma |c_0|^2|\psi_0|^2\psi_0\right),
\]
where $\rho=\mbox{sgn}(\mu-\mu_0)$. The solvability condition of
this $\psi_1$ equation is that its right hand side be orthogonal to
the adjoint homogeneous solution $\psi_0^*$. This condition yields
an equation for $c_0$ as
\begin{equation}  \label{f:c02} |c_0|^2 =
\frac{\rho \langle \psi_0^*, \psi_0\rangle}{\sigma  \langle
\psi_0^*, |\psi_0|^2\psi_0\rangle}.
\end{equation}

For the real eigenvalue $\mu_0$, its eigenfunction $\psi_0$
possesses partial \PT symmetry. Thus the two inner products in the
above equation are both real. Then for a certain sign of $\rho$,
i.e., when $\mu$ is on a certain side of $\mu_0$, the right side of
Eq. (\ref{f:c02}) is positive, hence this equation is solvable for
the constant $c_0$. Since the soliton in Eq. (\ref{e:soliton}) is
phase-invariant, we can take $c_0$ to be positive without any loss
of generality.

Pursuing this perturbation calculation to higher orders, we can find
that this perturbation solution can be constructed to all orders for
any small $\epsilon$, thus a continuous family of solitons
bifurcates out from the linear eigenmode $(\mu_0, \psi_0)$. In this
construction process, partial \PT symmetry of the potential is
critical. For instance, in the absence of this partial \PT symmetry
(and \PT symmetry), it is generally impossible to guarantee the
reality of inner products in Eq. (\ref{f:c02}), which makes this
equation unsolvable for $c_0$.

Next we corroborate these analytical results numerically. The
partial-\PT potential (\ref{e:Vexample}) with $\beta=0.1$ contains
three discrete real eigenvalues [see Fig. 1(c)]. From each of these
three eigenmodes, we have found numerically that a soliton family
bifurcates out, just as the theory predicted. To illustrate, we take
the focusing nonlinearity ($\sigma=1$). Then power curves of soliton
families bifurcated from the first and second eigenmodes of the
potential are displayed in Fig. 2. Here the power $P$ is defined as
$\int\int |\psi|^2dxdy$. Interestingly, these two power curves are
connected through a fold bifurcation, meaning that solitons
bifurcated from these two eigenmodes belong to the same solution
family, and the power of this solution family has an upper bound.
%This power-curve connectivity of solitons from different linear
%eigenmodes has also been reported in \cite{Zezyulin2012a} for a 1D
%harmonic \PT-symmetric potential, but that connectivity occurred
%under defocusing nonlinearity.

Profiles of solitons on this power curve are also displayed in Fig.
2. Here the amplitude fields of solitons at points `b,c' of the
power curve (with $\mu=1.3$) are plotted on the right column of the
figure. It is seen that the soliton at point `b' has higher
amplitude, obviously because it is on the upper power branch. The
phase fields of these two solitons are similar, thus only the phase
field at point `b' is shown. Note that these solitons share the same
partial \PT symmetry of the complex potential~(\ref{e:Vexample}).

\begin{figure}
\begin{center}
\includegraphics[height = 3in]{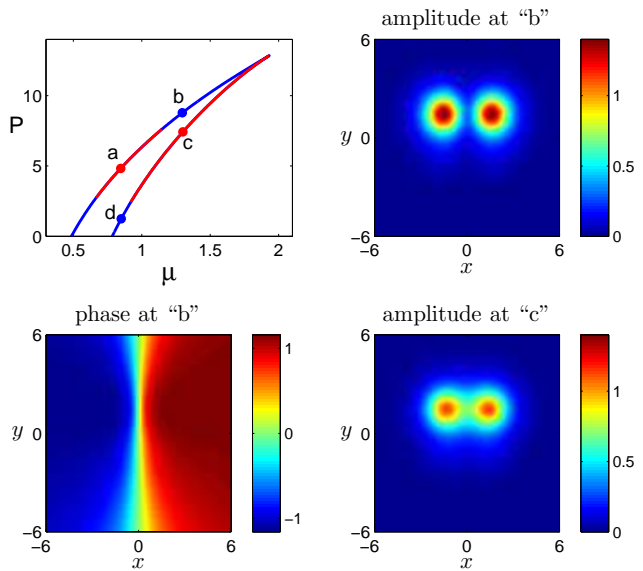}
\caption{Upper left: power diagram of the soliton family in potential (\ref{e:Vexample}) with $\beta=0.1$ and $\sigma=1$ (blue segments are stable and red unstable);
upper and lower right: amplitude fields of solitons ($|\psi|$) at points `b, c' of the power curve;
lower left: phase field of the soliton at point `b'. }
\end{center}
\end{figure}

Lastly, we examine linear stability of this soliton family. For this
purpose, we perturb these solitons by normal modes
\begin{equation*}
\Psi(x,y,z)=e^{i\mu z}\left[\psi(x,y)+f(x,y)e^{\lambda z}+g^*(x,y)e^{\lambda^* z}\right],
\end{equation*}
where $f, g\ll 1$, and $\lambda$ is the growth rate of the
disturbance. Linearization of Eq. (\ref{Eq:NLS}) for these
perturbations yields a linear-stability eigenvalue problem
\begin{equation}  \label{e:eigenvalue}
i\left[ {\begin{array}{cc}
   M_1 & M_2 \\
   -M_2^* & -M_1^* \\
         \end{array}}\right] \left[ {\begin{array}{c}
   {f} \\
   {g} \\
  \end{array}}\right] = \lambda  \left[ {\begin{array}{c}
   {f} \\
   {g} \\
  \end{array}}\right],
\end{equation}
where $M_1=\nabla^2+V-\mu +2\sigma |\psi|^2$, and $M_2=\sigma
\psi^2$. The soliton (\ref{e:solitonform}) is linearly unstable if
there exists an eigenvalue $\lambda$ such that $\mbox{Re}(\lambda) >
0$.

We solve this eigenvalue problem (\ref{e:eigenvalue}) by the Fourier
collocation method \cite{Yang_book}. For the four solitons on the
power curve of Fig. 2, their eigenvalue spectra are computed and
displayed in Fig. 3. It is seen that the soliton at point `a'
contains a quartet of complex eigenvalues, and the soliton at point
`c' contains a pair of real eigenvalues, thus these two solitons are
both linearly unstable. However, solitons at points `b,d' only
contain purely imaginary eigenvalues and are thus linearly stable.
%We note that the complex eigenvalues at point `a' are
%created when pairs of purely imaginary eigenvalues [see Fig. 3(b)]
%coalesce on the imaginary axis and then bifurcate off the imaginary
%axis, and the real eigenvalues at point `c' are generated when a
%pair of purely imaginary eigenvalues [see Fig. 3(d)] coalesce at the
%origin and then split off along the real axis.

Repeating this spectrum computation for other solitons on the power
curve of Fig. 2, their linear stability is then determined, and the
results are indicated on that power curve, with blue color
representing stable solitons and red color for unstable ones. Notice
that most of the lower power branch is unstable, while most of the
upper power branch is stable. This is surprising, since in
conservative potentials solitons on the upper power branch are
generally less stable. The increased stability of the upper power
branch here is clearly due to the complex partially-\PT-symmetric
potential (\ref{e:Vexample}), which stabilizes solitons at higher
powers.

\begin{figure}
\begin{center}
\includegraphics[height = 2.8in]{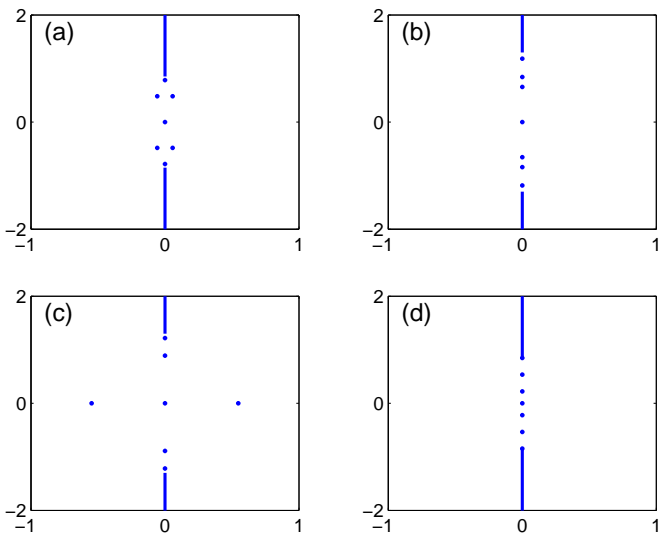}
\caption{Linear-stability spectra of the four solitons marked by letters `a,b,c,d' on the power curve of Fig.~2. }
\end{center}
\end{figure}

In summary, we have proposed a class of multi-dimensional complex
optical potentials that are not \PT-symmetric but rather
partially-\PT-symmetric, i.e., they are invariant under complex
conjugation and reflection in a single spatial direction. We have
shown that these partially-\PT-symmetric potentials can possess
all-real spectra and support continuous families of solitons,
similar to \PT-symmetric potentials. We have also shown that these
soliton families can exhibit multiple power branches, with the upper
power branches more stable than the lower ones. These results expand
the concept of \PT symmetry in multi-dimensions, and they may find
interesting optical applications.

The author thanks Prof. Vladimir Konotop for helpful discussions.
This work is supported in part by AFOSR and NSF.


\vspace{-0.2cm}
\begin{thebibliography}{99}
\newcommand{\enquote}[1]{``#1''}

\bibitem{Kivshar_book} Y.S. Kivshar and G.P. Agrawal, \emph{Optical Solitons: From Fibers to Photonic Crystals}
(Academic Press, San Diego, 2003).

\bibitem{Bender1998}
C.~Bender and S.~Boettcher,
%\enquote{{Real spectra in non-Hermitian Hamiltonians having PT symmetry},}
Phys. Rev. Lett. \textbf{80}, 5243--5246 (1998).

\bibitem{PT_2005}
A. Ruschhaupt, F. Delgado and J.~G. Muga,
%\enquote{Physical realization of PT-symmetric potential scattering in a planar slab waveguide},
J. Phys. A \textbf{38}, L171–-L176 (2005).

\bibitem{Christodoulides2007}
R.~El-Ganainy, K.~G. Makris, D.~N. Christodoulides and Z.~H.
Musslimani,
% \enquote{Theory of coupled optical PT-symmetric structures,}
Opt. Lett. \textbf{32}, 2632--2634 (2007).

\bibitem{Guo2009}
A.~Guo, G.~J. Salamo, D.~Duchesne, R.~Morandotti, M.~Volatier-Ravat,
V.~Aimez, G.~A. Siviloglou and D.~N. Christodoulides,
% \enquote{Observation of PT-Symmetry Breaking in Complex Optical Potentials,}
Phys. Rev. Lett. \textbf{103}, 093902 (2009).

\bibitem{Segev2010}
C.~E. Rueter, K.~G. Makris, R.~El-Ganainy, D.~N. Christodoulides,
M.~Segev and D.~Kip,
% \enquote{Observation of parity-time symmetry in optics,}
Nature Physics \textbf{6}, 192--195 (2010).

\bibitem{PT_lattice_exp}
A. Regensburger, C. Bersch, M.A. Miri, G. Onishchukov, D.N.
Christodoulides and U. Peschel,
%``Parity–time synthetic photonic lattices",
Nature \textbf{488}, 167-–171 (2012).

\bibitem{Musslimani2008}
Z.~H. Musslimani, K.~G. Makris, R.~El-Ganainy and D.~N.
Christodoulides,
% \enquote{Optical solitons in PT periodic potentials,}
Phys. Rev. Lett. \textbf{100}, 030402 (2008).

\bibitem{Wang2011} H. Wang and J. Wang,
%``Defect solitons in parity-time periodic potentials.
Opt. Exp. 19, 4030--4035 (2011).

\bibitem{Lu2011} Z. Lu and Z. Zhang,
%``Defect solitons in parity-time symmetric superlattices,"
Opt. Exp. 19, 11457--11462 (2011).

\bibitem{Nixon2012}
S.~Nixon, L.~Ge and J.~Yang,
%\enquote{Stability analysis for solitons in PT-symmetric optical lattices,}
Phys. Rev. A \textbf{85}, 023822 (2012).

\bibitem{Zezyulin2012a} D.~A. Zezyulin and V.~V. Konotop,
%``Nonlinear modes in the harmonic PT-symmetric potential,"
Phys. Rev. A 85, 043840 (2012).

\bibitem{Yang_necessity}
J. Yang,
%``Necessity of PT symmetry for soliton families in one-dimensional complex potentials",
arXiv:1310.4490 [nlin.PS] (2013) (to appear in Phys. Lett. A).

\bibitem{Longhi_2009} S. Longhi,
% \enquote{Bloch oscillations in complex crystals with PT symmetry},
Phys. Rev. Lett. \textbf{103}, 123601 (2009).

\bibitem{Musslimani_diffraction_2010}
K.~G. Makris, R.~El-Ganainy, D.~N. Christodoulides and Z.~H.
Musslimani,
%\enquote{PT-symmetric optical lattices},
Phys. Rev. A \textbf{81}, 063807 (2010).

\bibitem{Christodoulides_uni_2011}
Z. Lin, H. Ramezani, T. Eichelkraut, T. Kottos, H. Cao and D.N.
Christodoulides,
% \enquote{Unidirectional invisibility induced by PT-symmetric periodic structures},
Phys. Rev. Lett. \textbf{106}, 213901 (2011).

\bibitem{Abdullaev2011} F. K. Abdullaev, Y. V. Kartashov, V. V. Konotop and D. A.
Zezyulin,
%``Solitons in PT-symmetric nonlinear lattices."
Phys. Rev. A 83, 041805 (2011).

\bibitem{Li2011}
K. Li and P. G. Kevrekidis,
%``PT-symmetric oligomers: Analytical solutions, linear stability, and nonlinear dynamics,"
Phys. Rev. E 83, 066608 (2011).

\bibitem{coupler1}
R. Driben and B.A. Malomed,
%``Stability of solitons in parity-time-symmetric couplers",
Opt. Lett. 36, 4323 (2011).

\bibitem{He2012} Y. He, X. Zhu, D. Mihalache, J. Liu and Z. Chen,
%``Lattice solitons in PT-symmetric mixed linear-nonlinear optical lattices,"
Phys. Rev. A 85, 013831 (2012).

\bibitem{Zezyulin2012b} D. A. Zezyulin and V.V. Konotop,
%``Nonlinear Modes in Finite-Dimensional PT-Symmetric Systems,"
Phys. Rev. Lett. 108, 213906 (2012).

\bibitem{Nixon2012b}
S.~Nixon, Y.~Zhu, and J.~Yang,
%``Nonlinear dynamics of wave packets in PT-symmetric optical lattices near the phase transition point",
Opt. Lett. \textbf{37}, 4874-4876 (2012).

\bibitem{coupler2}
I.V. Barashenkov, S.V. Suchkov, A.A. Sukhorukov, S.V. Dmitriev, and
Y.S. Kivshar,
%``Breathers in PT-symmetric optical couplers",
Phys. Rev. A 86, 053809 (2012).

\bibitem{Konotop2012} V. V. Konotop, D. E. Pelinovsky and D. A. Zezyulin,
%``Discrete solitons in PT-symmetric lattices,"
Euro. Phys. Lett. 100, 56006 (2012).

\bibitem{Nixon2013}
S. Nixon and J. Yang,
%``Pyramid diffraction in parity-time-symmetric optical lattices",
Opt. Lett. 38, 1933--1935 (2013).

\bibitem{Kartashov2013}
Y.V. Kartashov,
%``Vector solitons in parity-time-symmetric lattices,"
Opt. Lett. 38, 2600--2603 (2013).

\bibitem{Barashenkov2013}
I. V. Barashenkov, L. Baker and N. V. Alexeeva,
%\PT-symmetry breaking in a necklace of coupled optical waveguides,
Phys. Rev. A 87, 033819 (2013).

\bibitem{Kevrekidis2013}
P. G. Kevrekidis,  D. E. Pelinovsky and D. Y. Tyugin,
%Nonlinear stationary states in PT-symmetric lattices,
SIAM J. Appl. Dyn. Syst., 12, 1210-1236 (2013).

\bibitem{Huang2013} C. Huang, C. Li, and L. Dong,
%``Stabilization of multipole-mode solitons in mixed linear-nonlinear lattices with a PT symmetry,"
Opt. Exp. 21, 3917--3925 (2013).

\bibitem{Yang_book} J. Yang, \emph{Nonlinear Waves in Integrable and Nonintegrable Systems} (SIAM, Philadelphia, 2010).

\end{thebibliography}
\end{document}